\documentclass{article}

\newcommand{\ii}{\mathrm{i}}
\newcommand{\ee}{\mathrm{e}}

\usepackage{amsmath}
\usepackage{graphicx}
\usepackage{psfrag}
\usepackage{amssymb}
\usepackage{color}

\usepackage{epstopdf, epsfig}
\usepackage{geometry}
\usepackage{amsmath}
\usepackage{subcaption}
\usepackage{xcolor}
\usepackage{bbm}

\newcommand{\andre}[1]{{{#1}}}

\begin{document}

\title{Reduced-order Galerkin models of plane Couette flow}%

\author{Andr\'e V. G. Cavalieri\footnote{Divis\~ao de Engenharia Aeroespacial, Instituto Tecnol\'ogico de Aeron\'autica, S\~ao Jos\'e dos Campos, SP, Brazil, andre@ita.br} \and Petr\^onio A. S. Nogueira\footnote{Department of Mechanical and Aerospace Engineering, Laboratory for Turbulence Research in Aerospace and Combustion, Monash University, Clayton, Australia, Petronio.Nogueira@monash.edu}}%
\date{\today}%
\maketitle

\begin{abstract}
Reduced-order models were derived for plane Couette flow using Galerkin projection, with orthonormal basis functions taken as the leading controllability modes of the linearised Navier-Stokes system for a few low wavenumbers. Resulting Galerkin systems comprise ordinary differential equations, with a number of degrees of freedom ranging from 144 to 600, which may be integrated to large times without sign of numerical instability. The reduced-order models so obtained are also found to match statistics of direct numerical simulations at Reynolds number 500 and 1200 with reasonable accuracy, despite a truncation of orders of magnitude in the degrees of freedom of the system. The present models offer thus an interesting compromise between simplicity and accuracy in a canonical wall-bounded flow, with relatively few modes representing coherent structures in the flow and their dominant dynamics.
\end{abstract}

The complexity of the Navier-Stokes equation has motivated the development of reduced-order models (ROMs) over the years. By reducing a full model to a simplified system, computational gains may be achieved, but, perhaps more importantly, one greatly simplifies the dynamics, allowing a closer inspection of the mechanisms at play~\cite{waleffe1997self,noack2003hierarchy}. ROMs are also important to derive flow control schemes~\cite{barbagallo2009closed}, as obtaining control laws may become prohibitively expensive when dealing with the full system.

A widespread technique for model reduction is Galerkin projection, where partial differential equations are projected onto a reduced set of modes that form a basis of a subspace of interest~\cite{rowley2017model}. This is also attractive as such modes may be related to coherent structures in flows~\cite{taira2017modal}. When Galerkin projections are applied to three-dimensional spatial modes, one obtains a system of ordinary differential equations that models the time evolution of mode amplitude~\cite{noack2003hierarchy}, and one may easily explore the interplay between coherent structures, or modes, by examining energy exchanges in the system~\cite{noack2005need,cavalieri2021structure}. In some cases, Fourier modes may be directly used in a Galerkin projection~\cite{saltzman1962finite,waleffe1997self}, but in more complex flows, with non-homogeneous directions, Fourier modes may not be applicable. A common choice of modes is based on proper orthogonal decomposition (POD) of numerical or experimental databases. POD provides an orthonormal basis that is optimal in representing the energy of the database at hand, and, for that property, was chosen in the derivation of a few ROMs in the literature~\cite{aubry1988dynamics,noack2003hierarchy,smith2005low,khoo2022sparse}.


Unlike Galerkin models built with analytical basis functions~\cite{waleffe1997self,moehlis2004low,cavalieri2021structure,cavalieri2022transition}, it is often mentioned that numerical Galerkin systems built using POD require a closure model to represent neglected modes, as discussed by Grimberg \emph{et al.}~\cite{grimberg2020stability} and Callaham \emph{et al.}~\cite{callaham2022role}; otherwise, one may obtain numerically unstable ROMs. Such closure models were developed in the early work of Aubry \emph{et al.}~\cite{aubry1988dynamics}, and modelling works have attempted various forms of closure in order to obtain a numerically stable and accurate Galerkin system~\cite{smith2005low,callaham2022role,khoo2022sparse}; a comprehensive review is provided by Ahmed \emph{et al.}~\cite{ahmed2021closures}. In this work we derive numerical Galerkin models for Couette flow with arbitrary order, which will be shown to be numerically stable and reproduce standard turbulence statistics with reasonable accuracy. This raises questions on the actual need of closure models to stabilise Galerkin models. Plane Couette flow is an ideal case for such a study, as a canonical flow for studies of transition and of wall-bounded turbulence. Successful model reductions of the Navier-Stokes system have already been obtained, either by a Galerkin projection along a spatial direction~\cite{lagha2007modeling}, or by restricting the possible non-linear interactions in the system~\cite{thomas2015minimal}.

Following the same conventions in \cite{cavalieri2021structure}, we consider incompressible plane Couette flow in Cartesian coordinates $(x,y,z)$ corresponding to streamwise, wall-normal and spanwise directions, with zero pressure gradient along $x$. Quantities are normalised by wall velocity and half wall separation, such that the walls are placed at $y=\pm 1$ and move with velocity $u=\pm 1$. The velocity vector $\mathbf{u}$, taken as fluctuations around the laminar solution $u_0 = y$, has components $u$, $v$ and $w$ along each Cartesian coordinate. We consider a computational domain with periodic boundary conditions in $x$ and $z$, with lengths given by $L_x$ and $L_z$, respectively. Non-slip boundary conditions are imposed on the walls at $y=\pm 1$.

Galerkin models are obtained based on an inner product defined as
\begin{equation}
\langle{\mathbf{f},\mathbf{g}} \rangle = \frac{1}{2L_x L_z}\int_0^{L_z}{\int_{-1}^{1}{\int_0^{L_x}{\mathbf{f}(x,y,z) \cdot \mathbf{g}(x,y,z) \mathrm{d}x}\mathrm{d}y}\mathrm{d}z},
\end{equation}
which for velocity fields in incompressible flows corresponds to the volume averaged kinetic energy.

We consider a modal decomposition of the velocity fluctuations as
\begin{equation}
\mathbf{u}(x,y,z,t) = \sum_j {a_j(t) \mathbf{u}_j(x,y,z)},
\label{eq:modaldecomposition}
\end{equation}
where each of the spatial modes $\mathbf{u}_j(x,y,z)$ satisfies the continuity equation and the boundary conditions, and form an orthonormal basis. We consider the Navier-Stokes equation for fluctuations around the laminar solution,
\begin{equation}
\frac{\partial \mathbf{u}}{\partial t} + (\mathbf{u}_0 \cdot \nabla) \mathbf{u} + (\mathbf{u} \cdot \nabla) \mathbf{u}_0 + (\mathbf{u} \cdot \nabla) \mathbf{u} = -\nabla p + \frac{1}{Re}\nabla^2 \mathbf{u},
\label{eq:navierstokes}
\end{equation}
where $p$ is the pressure. Inserting the decomposition of eq. (\ref{eq:modaldecomposition}) in eq. (\ref{eq:navierstokes}), and taking an inner product with $\mathbf{u}_i$, leads to a system of ordinary differential equations given by
\begin{equation}
\frac{\mathrm{d} a_i}{\mathrm{d} t}
=
\frac{1}{Re}\sum_j{L_{i,j}}a_j
+ \sum_j{\tilde{L}_{i,j}}a_j
+ \sum_j{\sum_k{Q_{i,j,k}}a_ja_k},
\label{eq:galerkinsys}
\end{equation}
where
\begin{equation}
L_{i,j} = \langle \nabla^2 \mathbf{u_j},\mathbf{u_i} \rangle,
\end{equation}
\begin{equation}
\tilde{L}_{i,j} = -\langle \left[(\mathbf{u_j} \cdot \nabla) \mathbf{u_0} + (\mathbf{u_0} \cdot \nabla) \mathbf{u_j} \right],\mathbf{u_i} \rangle,
\end{equation}
\begin{equation}
Q_{i,j,k} = -\langle (\mathbf{u_j} \cdot \nabla) \mathbf{u_k},\mathbf{u_i} \rangle.
\end{equation}
The pressure is eliminated from the equations using the divergence-free property of the modes. The coefficients in the ODE system of eq. (\ref{eq:galerkinsys}) include two linear terms $L_{i,j}$ and $\tilde{L}_{i,j}$, corresponding respectively to the viscous term and to interaction with the laminar solution $\mathbf{u}_0$. The quadratic term $Q_{i,j,k}$ represents non-linear interactions among modes.

Orthogonal modes $\mathbf{u}_i$ may be obtained as eigenfunctions of the controllability Gramian, considering the linearised Navier-Stokes system forced with white noise in space and time \cite{farrell1993stochastic,jovanovic2005componentwise}. We follow the formulation of Jovanovic \& Bamieh~\cite{jovanovic2005componentwise} and obtain orthonormal velocity modes considering as base flow the laminar solution for Couette flow. This is done for combinations of streamwise and spanwise wavenumbers $k_x$ and $k_z$, as indicated in table \ref{tab:modes}. A normal-mode Ansatz for controllability modes implies a dependence $\mathbf{u}_i(x,y,z) = \mathbf{\hat{u}}_i(y)\ee^{\ii (k_x x + k_z z)}$, which leads to complex-valued modes; we here deal with real modes by taking separate modes for the real and imaginary parts of $\mathbf{\hat{u}}_i(y)\ee^{\ii (k_x x + k_z z)}$, which leads to a pair of modes that only differ by a phase shift of $\pi/2$. The wavenumbers $k_x$ and $k_z$ are taken as integer multiples of the fundamental wavenumbers $\alpha=2\pi/L_x$ and $\beta=2\pi/L_z$, respectively, in order to match the periodicity imposed by the domain dimensions $L_x$ and $L_z$. Since modes with $(-k_x,-k_z)$ are simply complex conjugates of $(k_x, k_z)$, only positive $k_x$ is considered; for $k_x=0$ only $k_z \ge 0$ needs to be included in the basis. For $k_x=k_z=0$, corresponding to mean-flow modes, the linearised operator becomes singular and we need to resort to a different option. In this case, we have taken Stokes modes, eigenfunctions of viscous diffusion, as in earlier works~\cite{waleffe1997self}. The number of controllability modes in table~\ref{tab:modes} was chosen to ensure the stability of the laminar solution of Couette flow; stronger truncations were seen to lead to an unstable laminar solution, which would not be physical.

\begin{table}[h!]
  \begin{center}
  \begin{tabular}{|c|c|c|c|c|c|c|c|c|c|}
  \hline
	$\mathrm{N}$ & $k_x/\alpha$ & $n_{modes,y}$ & $k_z/\beta$ \\ \hline
	144 & $0, 1$ & 16 & $-1, 0, 1$ \\ \hline
	360 & $ 0, 1$ & 24 & $-2, -1, 0, 1, 2$ \\ \hline
	600 & $0, 1, 2$ & 24 & $-2, -1, 0, 1, 2$ \\ \hline
  \end{tabular}
  \caption{Galerkin systems of this work}
\label{tab:modes}
  \end{center}
\end{table}

The choice of controllability modes is similar to the use of POD to derive several reduced-order models in the literature~\cite{aubry1988dynamics,noack2003hierarchy,khoo2022sparse}, as controllability modes correspond to POD for the linear system driven with white noise~\cite{farrell1993stochastic,jovanovic2005componentwise,bagheri2009input}. The advantage of using controllability modes is that one does not need a full numerical or experimental database to obtain POD modes; here, the linearised operator is used to obtain an orthonormal basis that approximates large-scale structures in the flow. \andre{Unless otherwise specified,} for the present ROMs we have constructed bases with controllability modes taken at a low Reynolds number, $\mathrm{Re}=100$. Values of Reynolds number matching the cases explored here, $\mathrm{Re}=500$ and $\mathrm{Re}=1200$, were also attempted, but led to slightly worse quantitative agreement with reference statistics; \andre{we will later illustrate some results for a basis constructed using $\mathrm{Re}=500$. The better performance of a basis constructed at low $\mathrm{Re}$} may be a further indication that linearised models obtained considering additional viscous effects, akin to an eddy viscosity, may be a better representation of coherent structures in turbulent flows, as explored in recent works~\cite{morra2019relevance,pickering2021optimal}. It is nonetheless interesting to use a fixed set of modes for various Reynolds numbers, as it allows using a single ROM for several $\mathrm{Re}$, such that invariant solutions and bifurcations may be studied with ease~\cite{moehlis2005periodic,cavalieri2022transition}.

\begin{figure}[h!]
\includegraphics[width=1.0\textwidth]{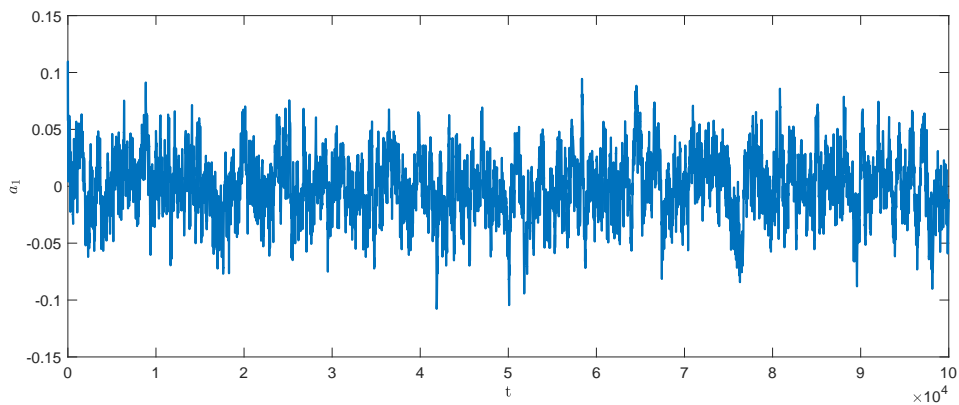}
\caption{Time series of the first mode coefficient, $a_1(t)$, from an integration of the $N=144$ ROM at $\mathrm{Re}=1200$.}
\label{fig:a1}
\end{figure}

The controllability modes are obtained using spectral methods, with a Chebyshev discretisation in $y$ following Weideman \& Reddy~\cite{weideman2000matlab}. The modes so obtained were verified to satisfy the continuity equation to machine precision. Quadratures were carried out using Clenshaw-Curtis quadrature in $y$~\cite{trefethen2000spectral}, and the standard trapezoidal rule in $x$ and $z$ as it has spectral accuracy for a Fourier discretisation~\cite{trefethen2014exponentially}. The domain considered here has lengths $(L_x,L_z)=(2\pi,\pi)$, a domain size often used in studies of transitional Couette flow~\cite{kreilos2014increasing}, discretised with 16 points in $x$ and $z$ and 65 points in $y$; the lower number of points in $x$ and $z$ is nonetheless well beyond the required amount for dealiasing the few Fourier modes at hand. Time integration is carried out with a standard 4th/5th Runge-Kutta method for 3000 non-dimensional time units in order to collect statistics. A random initial condition was taken, and statistics are collected after a transient of 500 time units. Albeit not shown here, the models display transient chaos, as in our earlier work~\cite{cavalieri2021structure}, but with longer lifetimes, well beyond the 3000 time units considered for statistics. The models have shown to be numerically stable, with integrations carried out to much higher times, up to $10^6$, which is much longer than typical direct numerical simulations of turbulent Couette flow. A sample time series in shown in fig. \ref{fig:a1}, with integration carried out for $10^5$ time units without any sign of numerical instability, such as reported in the literature to occur in about 100 non-dimensional time units~\cite{grimberg2020stability}. 

\begin{figure}[h!]
\begin{subfigure}{0.48\textwidth}
\includegraphics[width=1.0\textwidth]{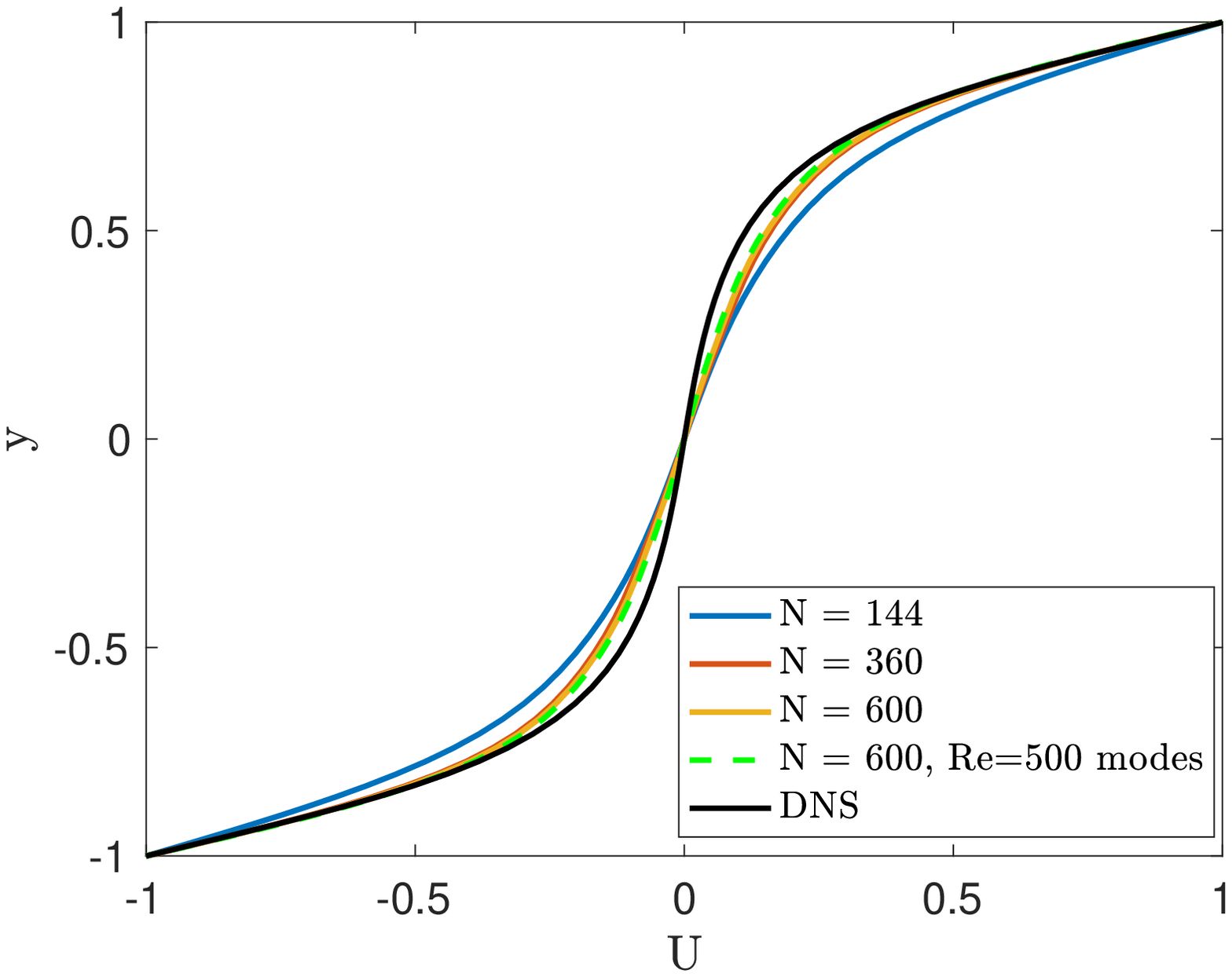}
\caption{Mean flow}
\end{subfigure}
\begin{subfigure}{0.48\textwidth}
\includegraphics[width=1.0\textwidth]{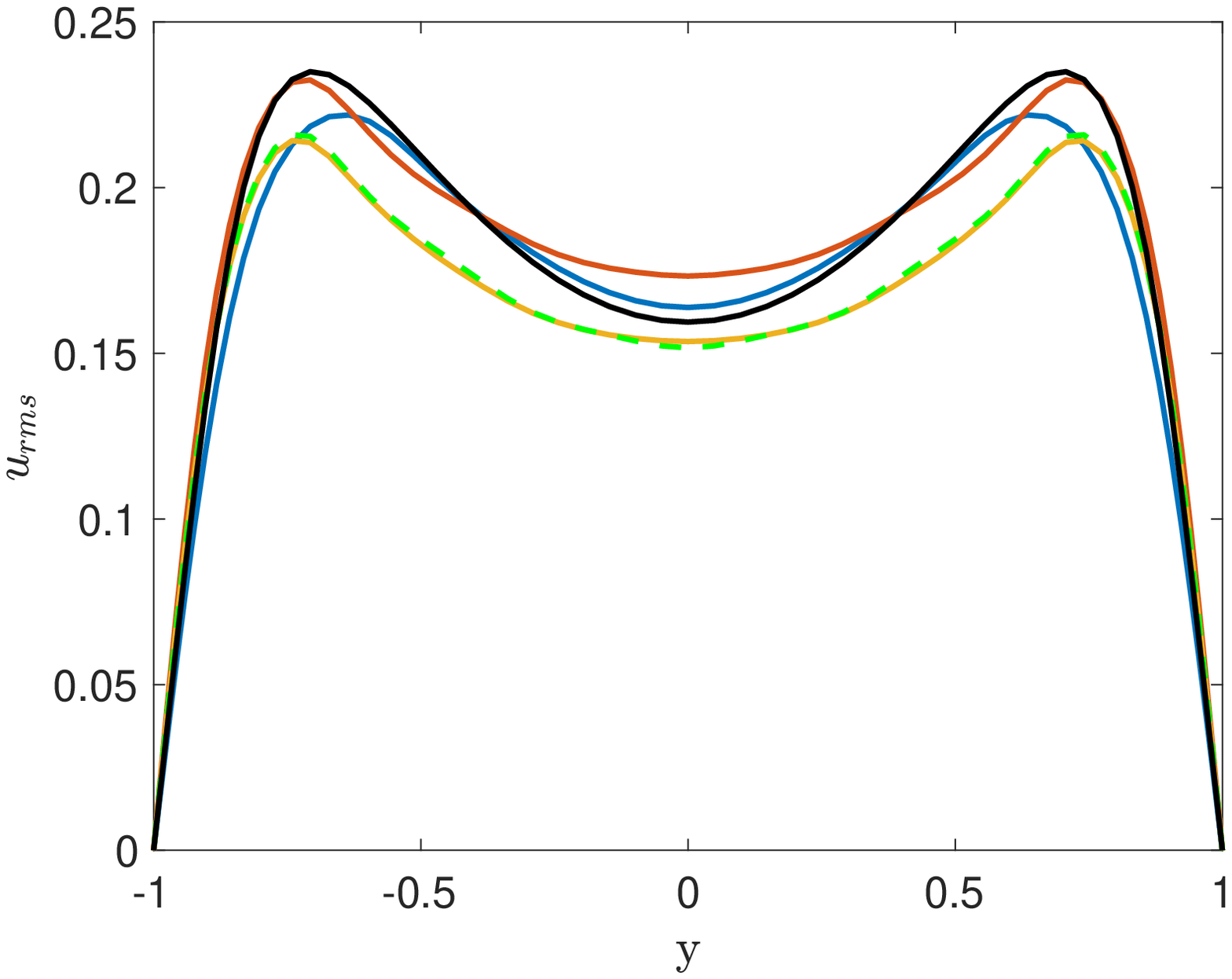}
\caption{Streamwise velocity fluctuations}
\end{subfigure}
\begin{subfigure}{0.48\textwidth}
\includegraphics[width=1.0\textwidth]{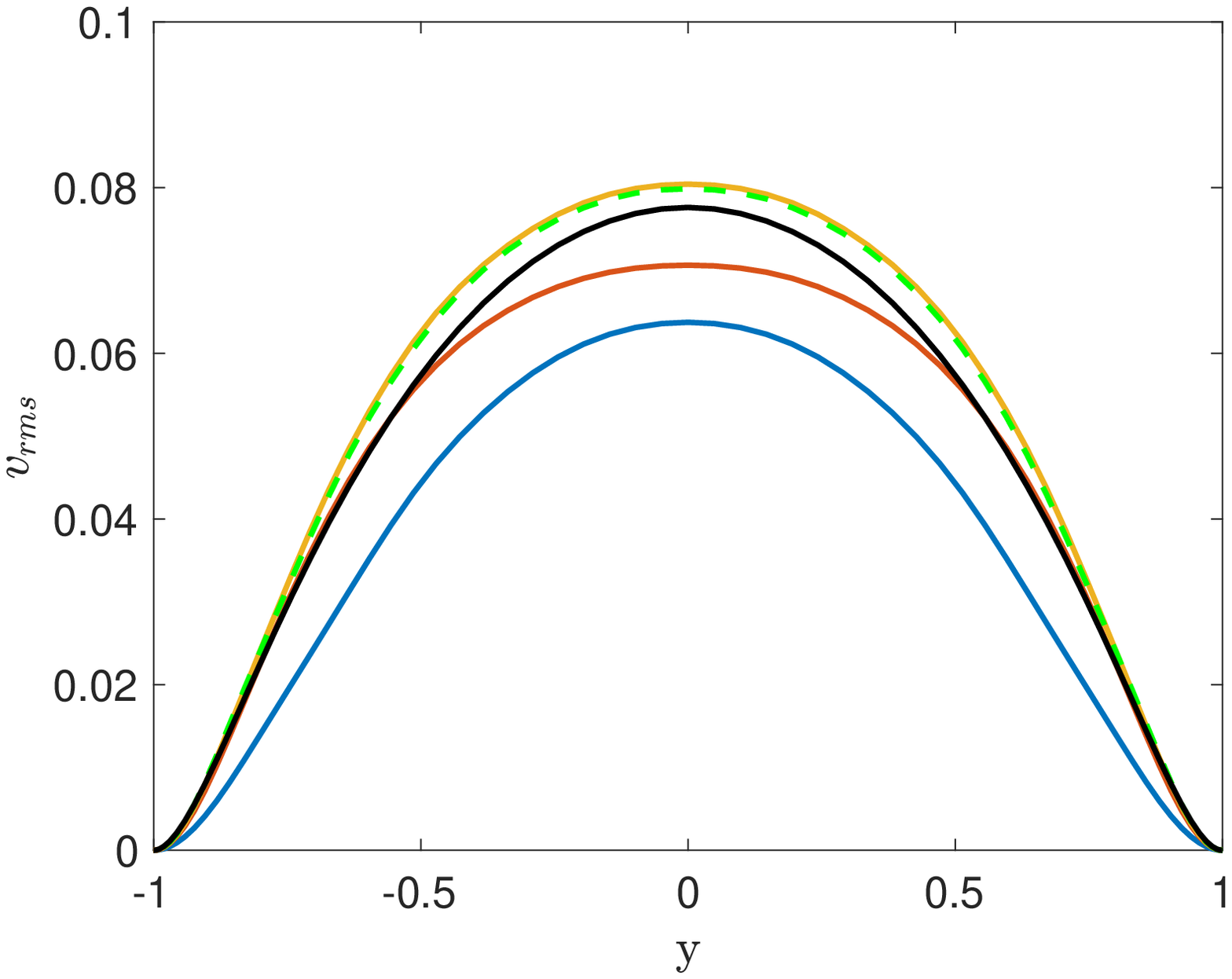}
\caption{Wall-normal velocity fluctuations}
\end{subfigure}
\begin{subfigure}{0.48\textwidth}
\includegraphics[width=1.0\textwidth]{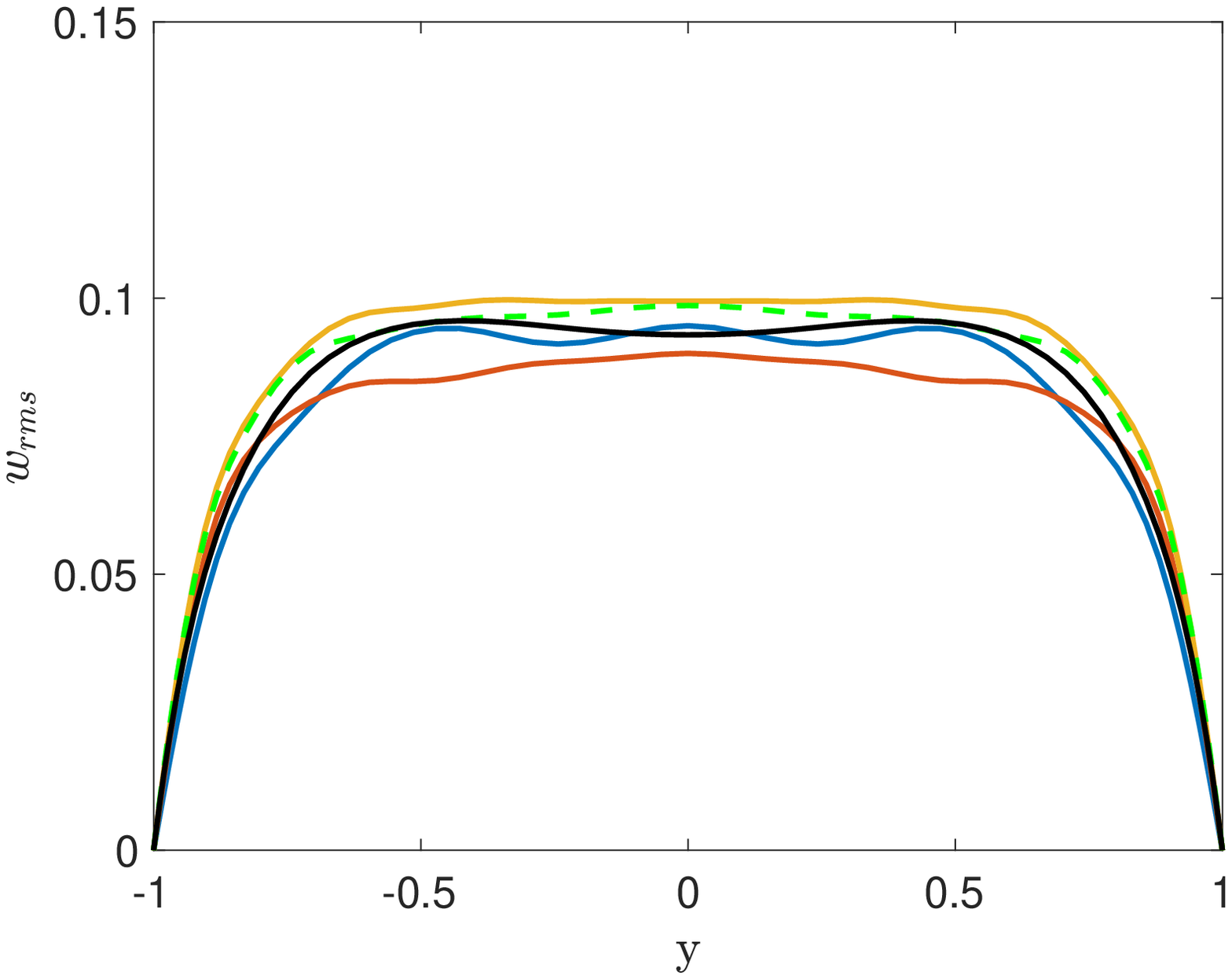}
\caption{Spanwise velocity fluctuations}
\end{subfigure}
\caption{Statistics of ROMs and DNS for $\mathrm{Re}=500$. \andre{Full coloured lines show results obtained with $\mathrm{Re}=100$ controllability modes; green dashed line for $\mathrm{Re}=500$ controllability modes}.}
\label{fig:Re500stats}
\end{figure}

The statistics obtained with the Galerkin systems of table \ref{tab:modes} are shown in figure \ref{fig:Re500stats} for $\mathrm{Re}=500$. We show the mean velocity profile $U$, which results from the $k_x=k_z=0$ modes in the ROM, and the root-mean-square (RMS) values of the three velocity components $u$, $v$ and $w$. Results are compared to reference DNS statistics obtained with Channelflow~\cite{gibson2019channelflow}, with a grid resolution of $(N_x,N_y,N_z) = (32,65,32)$, times a 3/2 factor in $N_x$ and $N_z$ for dealiasing; the DNS thus has a number of degrees of freedom (DoFs) of about $10^5$. \andre{We initially focus on the results obtained with $\mathrm{Re}=100$ controllability modes, displayed in full coloured lines.} The various ROMs display a reasonable agreement with the DNS statistics, with a somewhat lower accuracy for the strongest truncation with $N=144$ DoFs. The present reductions of 2 to 3 orders of magnitude retain nonetheless the salient features of turbulent Couette flow. For the truncations at hand, which are coarse approximations of the full system, we are still far from a monotonic convergence to DNS. Such non-monotonic convergence was observed for turbulent channel flow solved with coarse numerical resolutions~\cite{meyers2007plane,rasam2011effects}. One would thus expect monotonic convergence to reference statistics only for a number of DoFs significantly higher than the present models; however, it is encouraging that all present discretisations lead to statistics in fair agreement with DNS data. The present ROM displays an agreement to reference DNS results that is equal or better than earlier POD-Galerkin ROMs for plane Couette flow~\cite{smith2005low,khoo2022sparse}, which use DNS data to calibrate closure models. Here, no calibration is performed, and accurate dynamics are obtained by simply increasing the number of modes in the basis.

\begin{figure}[h!]
\begin{subfigure}{0.48\textwidth}
\includegraphics[width=1.0\textwidth]{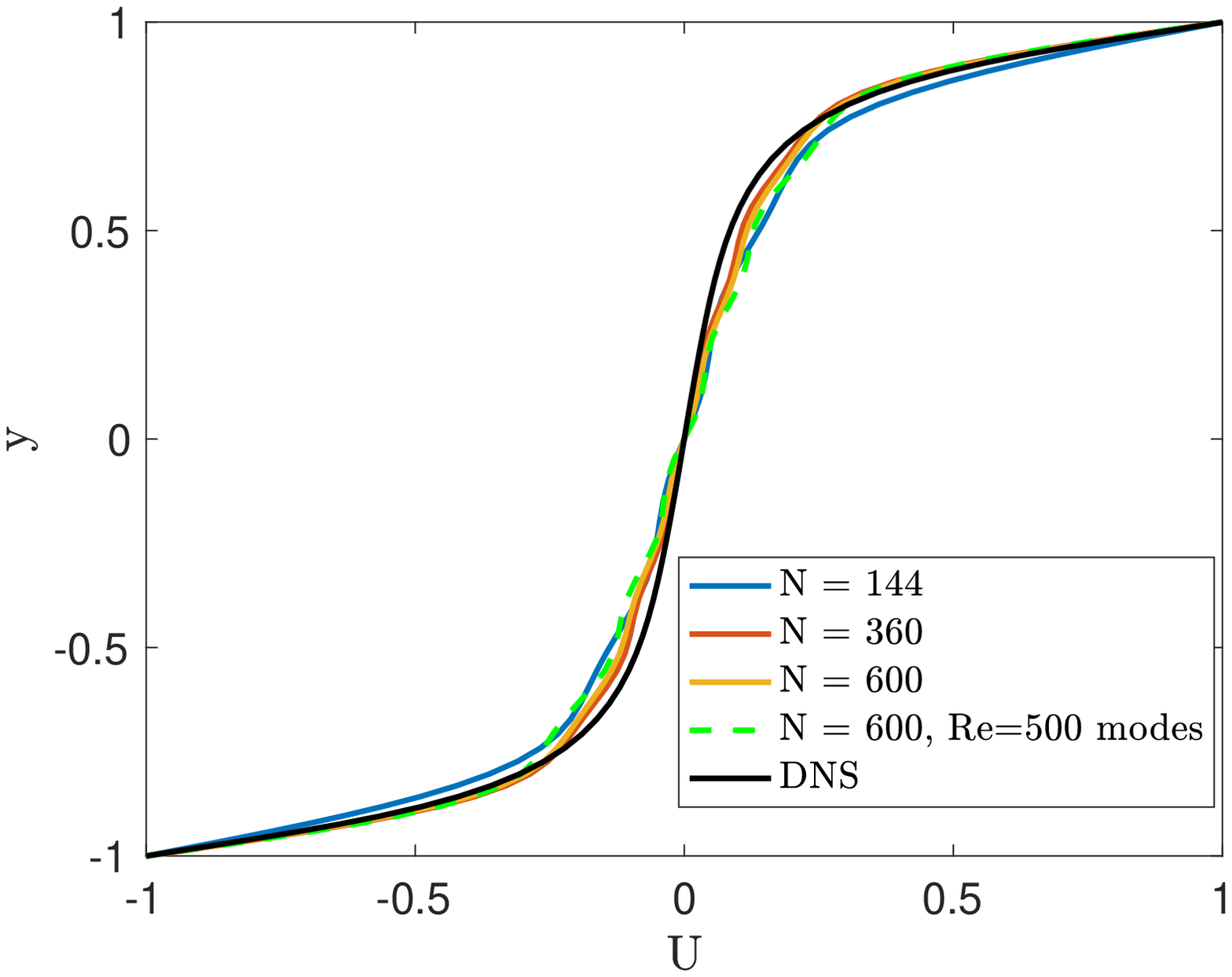}
\caption{Mean flow}
\end{subfigure}
\begin{subfigure}{0.48\textwidth}
\includegraphics[width=1.0\textwidth]{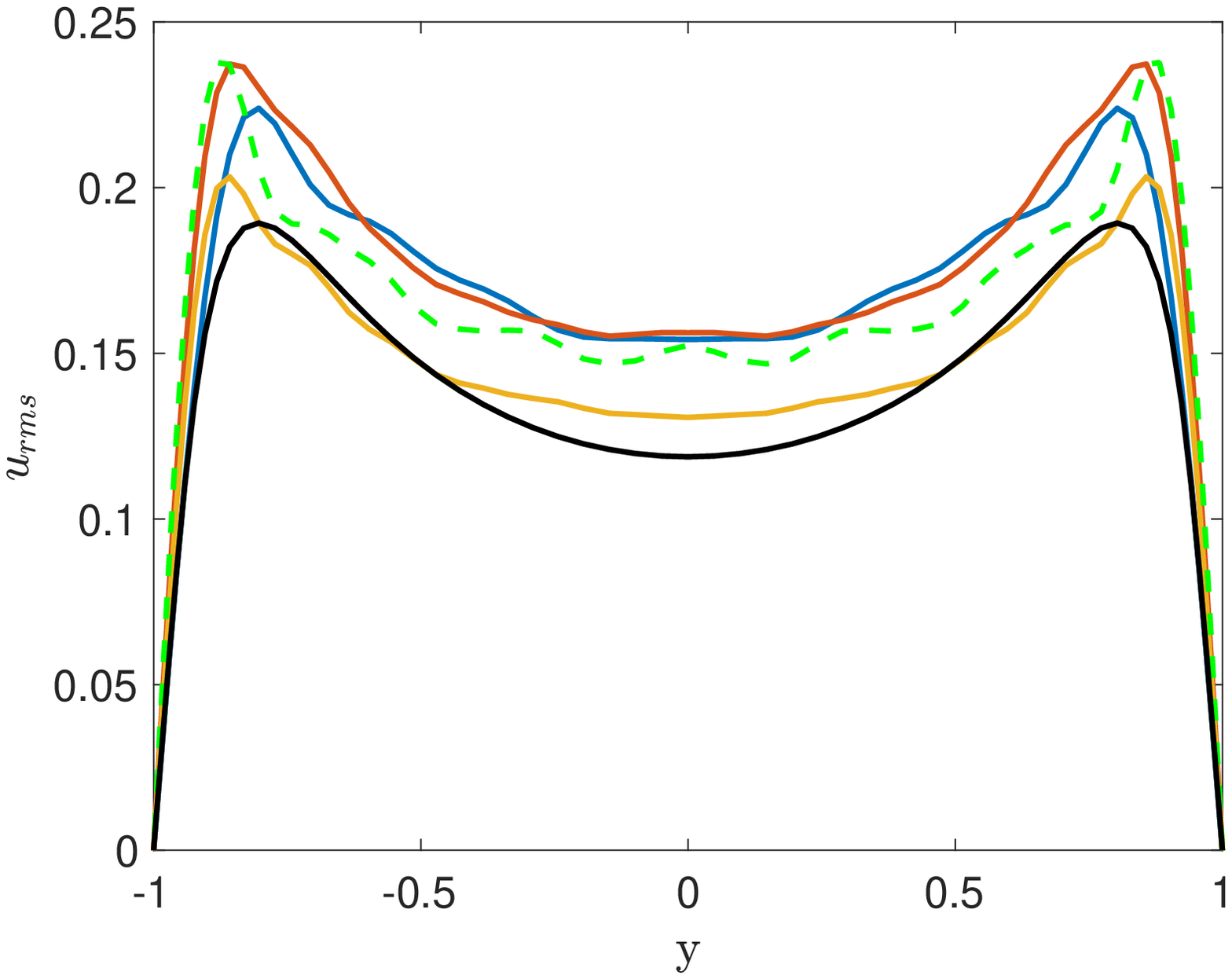}
\caption{Streamwise velocity fluctuations}
\end{subfigure}
\begin{subfigure}{0.48\textwidth}
\includegraphics[width=1.0\textwidth]{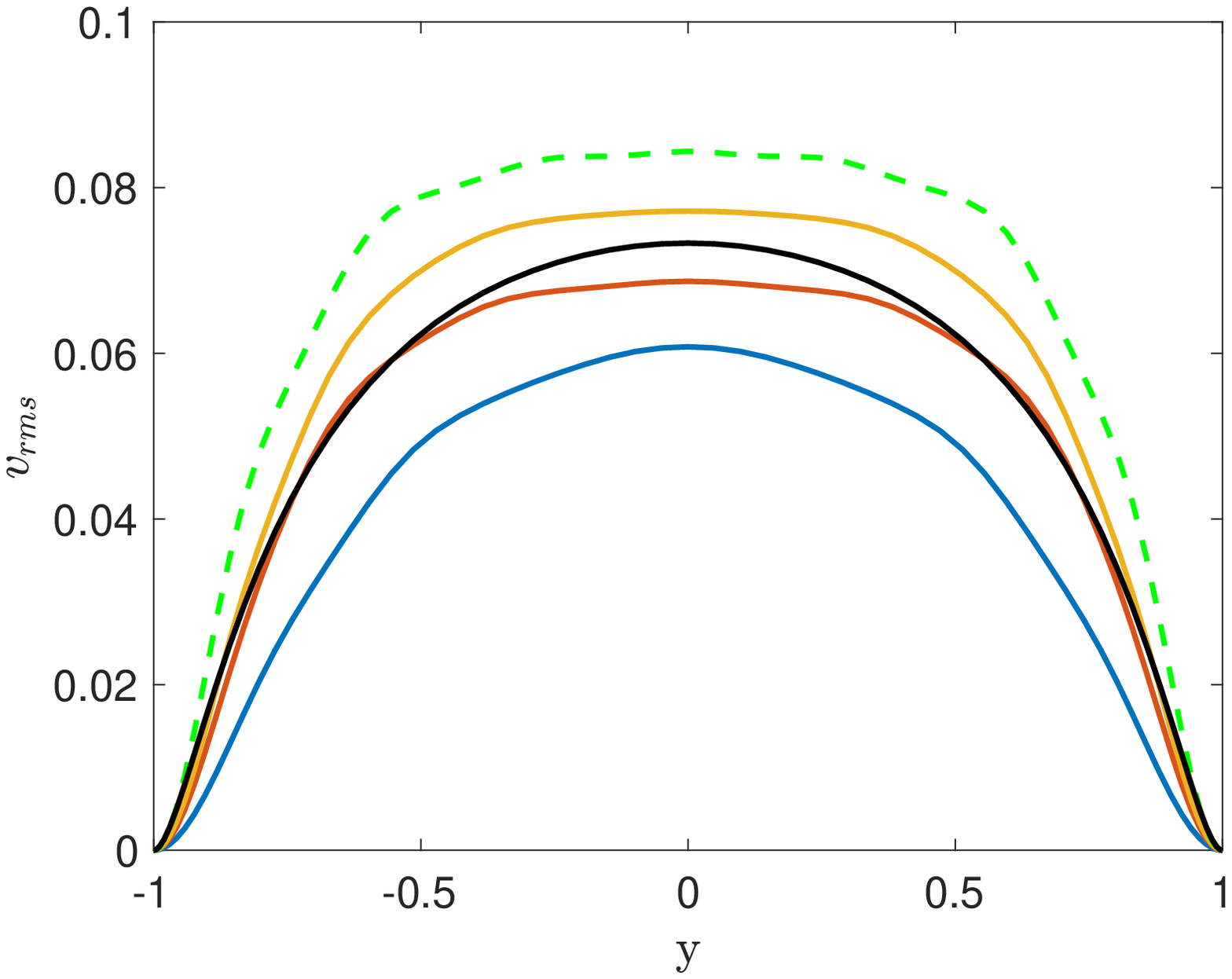}
\caption{Wall-normal velocity fluctuations}
\end{subfigure}
\begin{subfigure}{0.48\textwidth}
\includegraphics[width=1.0\textwidth]{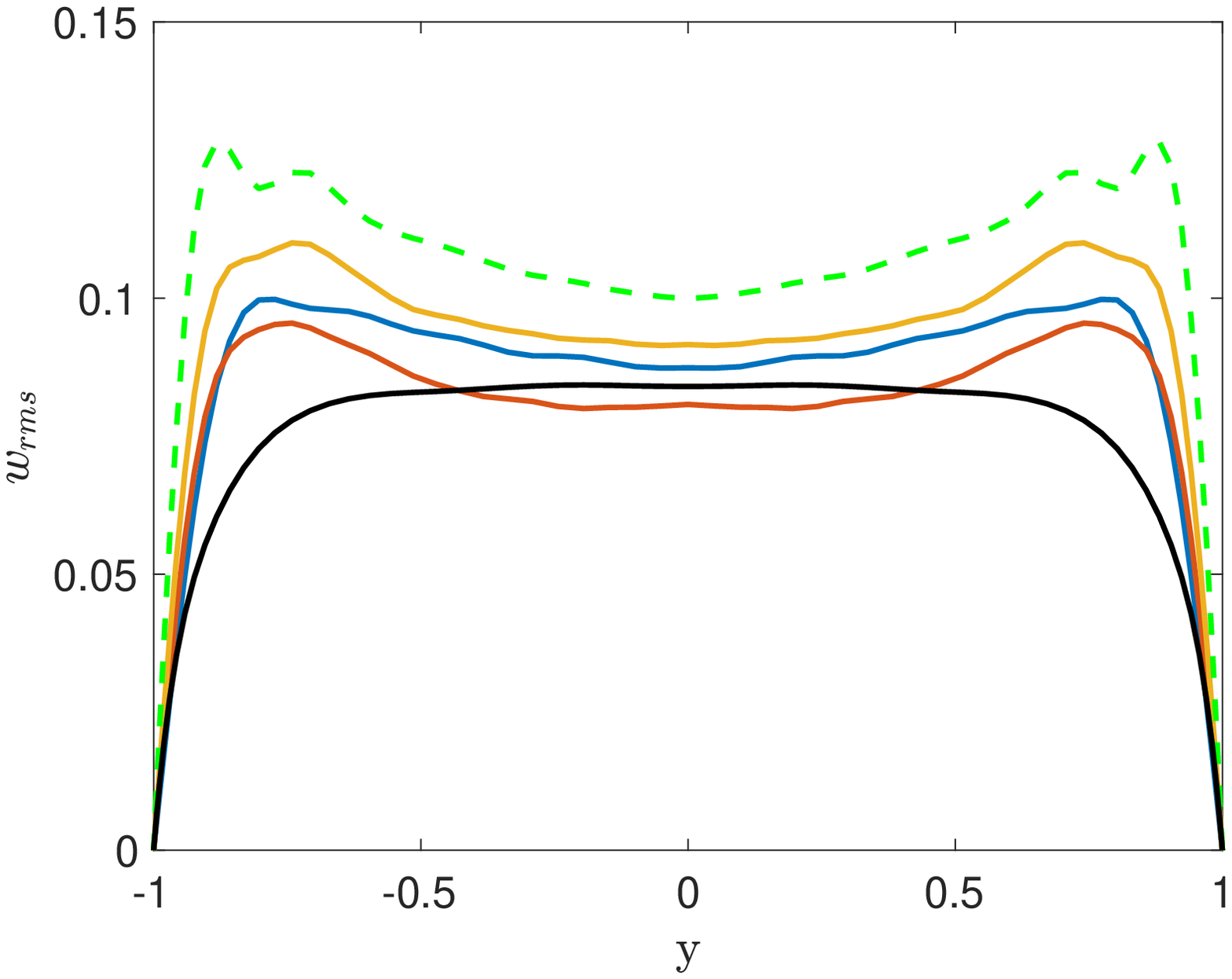}
\caption{Spanwise velocity fluctuations}
\end{subfigure}
\caption{Statistics of ROMs and DNS for $\mathrm{Re}=1200$. \andre{See caption of fig. \ref{fig:Re500stats}}.}
\label{fig:Re1200stats}
\end{figure}

The model results are compared to DNS for $\mathrm{Re}=1200$ in figure \ref{fig:Re1200stats}. In this case, the DNS was carried out with $(N_x,N_y,N_z) = (64,65,64)$ to maintain typical resolutions of direct numerical simulation, leading to a number of DoFs of about $5\cdot 10^5$. The present ROMs thus have a reduction of 3 to 4 orders of magnitude in the number of DoFs, and maintain nonetheless a reasonable quantitative agreement with reference statistics. The profiles in $y$ display oscillations in some of the cases, which we attribute to a low number of controllability modes to discretise the $y$ direction. This is observed for the $w$ statistics of the $N=144$ ROM for $\mathrm{Re}=500$, and for the three ROMs as $\mathrm{Re}$ is increased to 1200 in figure \ref{fig:Re1200stats}.

\andre{The green dashed lines in figures \ref{fig:Re500stats} and \ref{fig:Re1200stats} show statistics of the largest ROM, constructed with controlability modes taken at $\mathrm{Re}=500$. For the ``design'' $\mathrm{Re}=500$ this ROM has almost the same velocity statistics than the one constructed with $\mathrm{Re}=100$ controllability modes. This is reassuring, as it shows that the performance of the present models does not depend crucially on a specific choice of basis functions. However, when the basis is ``off-design'' at $\mathrm{Re}=1200$ the model statistics are further from the reference results. All ROMs tested are nonetheless stable and lead to reasonable statistics, but we observe that the convergence to DNS results depends on the choice of basis functions. In what follows we will restrict ourselves to the ROMs constructed with $\mathrm{Re}=100$ controllability modes.}

\begin{figure}[h!]
\begin{subfigure}{0.48\textwidth}
\includegraphics[width=1.0\textwidth]{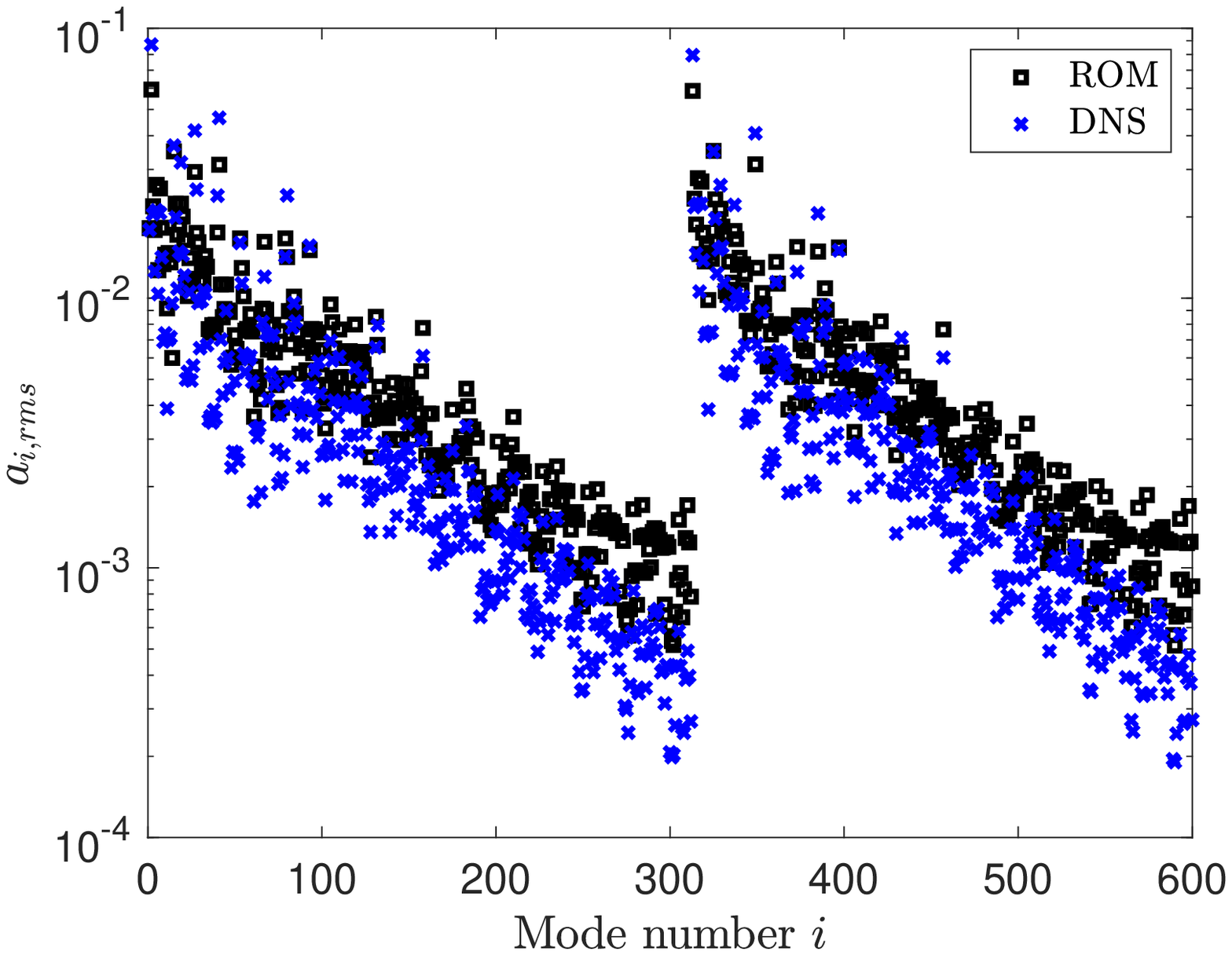}
\caption{$\mathrm{Re}=500$}
\end{subfigure}
\begin{subfigure}{0.46\textwidth}
\includegraphics[width=1.0\textwidth]{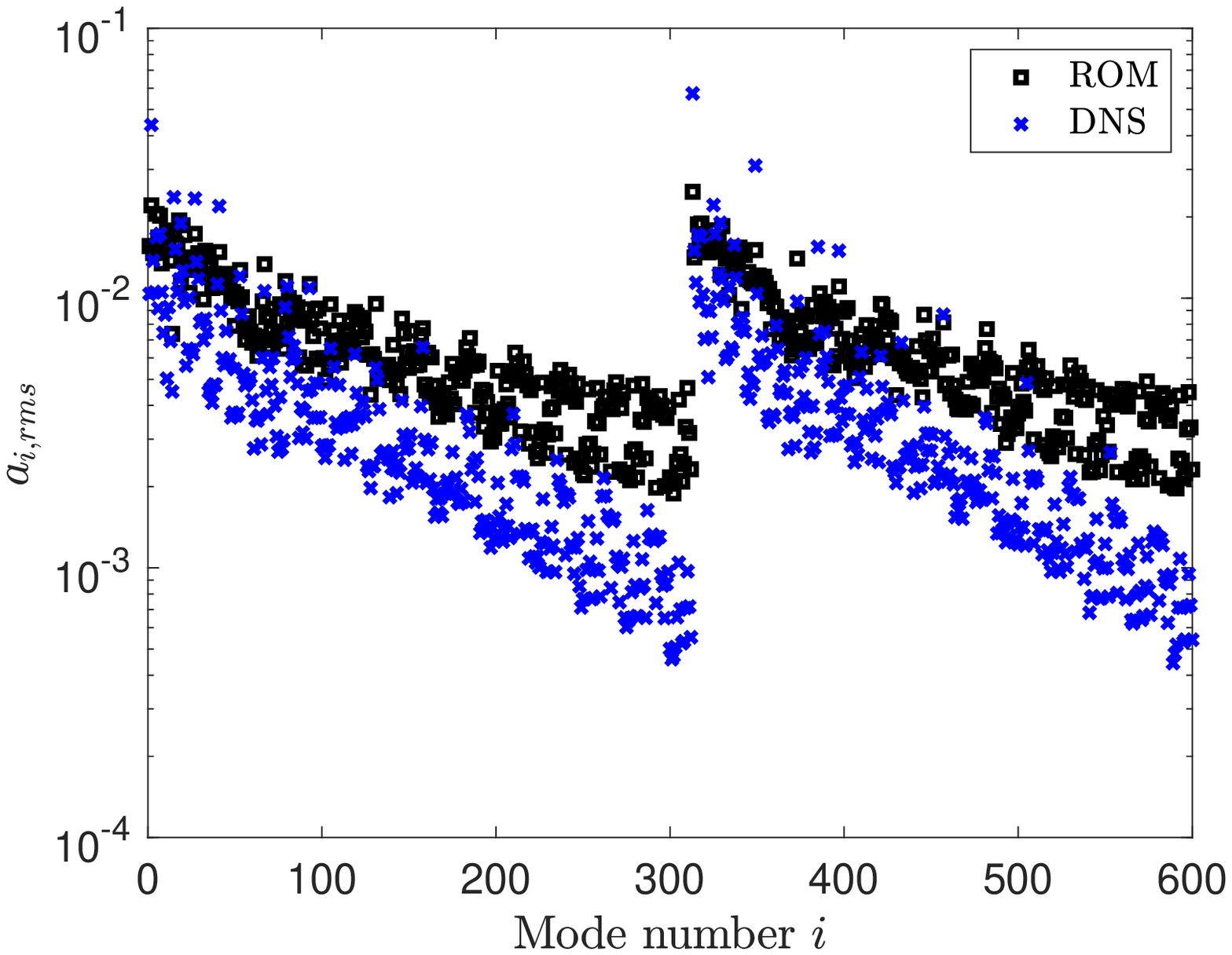}
\caption{$\mathrm{Re}=1200$}
\end{subfigure}
\caption{Root-mean-square values of temporal coefficients $a_i(t)$ obtained either from a ROM simulation or from projection of DNS data onto the ROM modal basis. Results for the $N=600$ ROM.}
\label{fig:rmsmodes}
\end{figure}

A further check of the accuracy of the ROMs was made by projection the DNS data onto the modal basis $\mathbf{u}_i$ as
\begin{equation}
a_{i}^{(DNS)}(t) = \langle \mathbf{u}_{DNS}, \mathbf{u}_i \rangle.
\label{eq:DNSprojection}
\end{equation}
and then computing the RMS values of time coefficients $a_i$, obtained either from the ROM or from the projection of DNS data onto the modal basis following eq. (\ref{eq:DNSprojection}). Results are shown in figure \ref{fig:rmsmodes} for the largest ROM with $N=600$. The projections show the same pattern for low and high mode number, as the higher modes are equal to the lower ones, but phase shifted by $\pi/2$. The results for both Reynolds numbers show RMS values of modes decaying with increasing mode number, as expected as one goes from larger to smaller structures in a cascade. Focusing on the results for $\mathrm{Re}=500$, we observe similar trends between ROM results and DNS projections, which indicates that the ROM accurately represents the dynamics of Couette flow in the subspace spanned by the modal basis. However, the RMS values in the ROM are generally higher than the DNS projections, which may be understood by the lower number of DoFs in the ROM. In the DNS the energy may be transferred to smaller structures in higher wavenumbers, a process that is halted in the ROM as one truncates the basis, a phenomenon known as spectral blocking~\cite{boyd2001chebyshev}. This is more visible in the $\mathrm{Re}=1200$ results, where RMS values of the ROM are significantly higher than the DNS projections, especially for higher-order modes, which is a symptom of accumulation of energy at higher modes due to their inability to transfer such energy to even smaller structures.

In summary, the ROMs obtained in this work, with a few hundred degrees of freedom, have shown to represent the statistics and dynamics of plane Couette flow with moderate Reynolds number. The various truncations explored here all led to Galerkin systems which may be integrated in time without numerical instabilities, with model results that are reasonably close to reference DNS data despite the strong truncations. This indicates that inclusion of closure models~\cite{ahmed2021closures,callaham2022role} is not an essential feature of Galerkin models that are robust and accurate. No numerical instability was observed in any of the present models, and a possible reason for this was the use of the same numerics of DNS in each step in the derivation of the ROM: the same pseudo-spectral methods of the DNS were used to compute derivatives and integrals. The robustness of the present model is similar to other Galerkin models built with analytical basis functions, which could be extensively integrated to large times without reported numerical issues~\cite{moehlis2004low,chantry2016turbulent,cavalieri2021structure}. However, inclusion of closure models may be beneficial for model accuracy, similarly to the use of subgrid models in large-eddy simulations. The use of plane Couette flow, with its simple geometry and nonetheless complex dynamics, plenty of available literature on transition and turbulence, and the availability of open-source codes for direct numerical simulation, is suggested as a benchmark case for further development of reduced-order models.

The present ROMs open a number of interesting directions for further work on turbulence dynamics. It is straightforward to include or remove modes of interest, or even specific non-linear interactions, and inspect the effect on the dynamics, as done in our previous work~\cite{cavalieri2021structure}. This may help to probe further wall-bounded turbulence in search of non-linear interactions with a dominant effect on dynamics. The observation that all ROMs in this work reproduce, within some error tolerance, the expected statistics of turbulent Couette flow, indicates that the bulk of the dynamics may be described by non-linear interaction involving a few modes representing coherent structures. Non-linear interactions involving streamwise wavenumbers have successfully been truncated in recent works~\cite{thomas2015minimal,bretheim2015standard,hernandez2022generalised}, and the present ROMs also involve truncations in wall-normal and spanwise directions, retaining nonetheless a good amount of the underlying dynamics. It is of course clear that in order to obtain DNS-type agreement with reference quantities one needs DNS-range numbers of degrees of freedom, but significantly reduced systems may nonetheless provide an interesting compromise between simplicity and fidelity. 

\section*{Acknowledgments}

This work was supported by FAPESP grant no. 2019/27655-3 and CNPq grant no. 313225/2020-6. The first author would like to thank the Isaac Newton Institute for Mathematical Sciences, Cambridge, for support and hospitality during the programme "Mathematical aspects of turbulence: where do we stand?" where work on this paper was undertaken. This work was supported by EPSRC grant no. EP/R014604/1 and by a grant from the Simons Foundation. The second author was supported by the Australian Research Council through the Discovery Project scheme: DP190102220. Numerical implementations of the present ROMs may be provided by the first author upon reasonable request.

\bibliographystyle{abbrv}
\bibliography{/Users/andrecavalieri/Desktop/LinuxData/Biblio/biblio}

\end{document}